# Collective scattering in hybrid nanostructures with many atomic oscillators coupled to an electromagnetic resonance

Pierre Fauché,[1,2] Spyridon G. Kosionis[1] and Philippe Lalanne[1,*]
[1] Laboratoire Photonique, Numérique et Nanosciences (LP2N), UMR 5298, CNRS-IOGS-Univ. Bordeaux, Institut d'Optique d'Aquitaine, 33400 Talence, France.
[2] CNRS, University Bordeaux, CRPP, UPR 8641, 115 Avenue Schweitzer, 33600 Pessac, France.
*Corresponding author: philippe.lalanne@institutoptique.fr

**Abstract.** There is considerable interest in collective effects in hybrid systems formed by molecular or atomic ensembles strongly coupled by an electromagnetic resonance. For analyzing such collective effects, we develop an efficient and general theoretical formalism based on the natural modes of the resonator. The main strength of our approach is its generality and the high level of analyticity enabled by modal analysis, which allows one to model complex hybrid systems without any restriction on the resonator shapes or material properties, and to perform statistical computations to predict general properties that are robust to spatial and polarization disorders. Most notably, we establish that superradiant modes remain even after ensemble averaging and act as an "invisibility cloak" with a spectral bandwidth that scales with the number of oscillators and the spatially-averaged Purcell factor.

## INTRODUCTION

Cooperative scattering by large collections of quantum emitters has become increasingly important in modern science and technology,[1-3] an emblematic example being Dicke superradiance.[4] Of particular interest are high cooperativity regimes, which are promoted by long-range interactions and quantum cavity-electrodynamic effects with electromagnetic resonances. Recent exemplary advances encompass the generation of coherent visible radiation by many emitters placed near plasmonic nanoparticles[5-6] or hybrid quantum systems combining cold-atom clouds with photonic-crystal resonances.[7-8]

Even for steady-state cases, the theoretical analysis of quantum hybrids represents a major challenge in computational electrodynamics, requiring the repeated calculation of the full-photon Green's functions of the resonator for different frequencies and atomic positions.[9-11] The challenge worsens when studying the dynamics by iteratively solving coupled equations for the Maxwell's fields and carrier-population operators,[12-13] or when computing ensemble-averaged responses[14-15] to interpret experiments for which the exact location and orientation of atoms are unknown.[16-18]

Here, we provide a powerful computational "toolkit" based on the natural modes of the resonator for analyzing collective effects in large ensembles of atoms or molecules coupled by an electromagnetic resonator, which could be a high-Q or low-Q micro-nanoresonator, possibly with a spectral overlap of several resonances. The main strength of our approach is the high level of analyticity brought on by the modal analysis of the emitter-resonator interaction, which enables one to perform statistical treatments and to predict general properties that are robust to spatial and polarization disorders. We adopt a classical polarizability model for the quantum emitters. Classical-oscillator treatments cannot describe all aspects of quantum hybrids, but in return provide a rudimentary and intuitive tool, which is general enough to predict many important features of quantum systems and is reusable in quantum formalisms.[19-22,10] Additionally, cooperativity in ordered or disordered collections of classical electromagnetic resonators is a topic of great interest on its own, for instance to synthesize complex spectral Fano responses[23] or create new hybrid resonances.[24]

**SCATTERING FORMALISM: CROSS SECTIONS**

We consider a collection of $N$ oscillators, $j = 1, 2 \ldots N$, located at positions $r_j$ near a nanostructure supporting a localized resonance. A typical setup would involve, e.g., dye molecules[5] or quantum dots[25] attached to a metal nanoparticle, or atoms trapped close to a dielectric photonic-crystal cavity.[7] The hybrid system is illuminated by an incident beam with an electric field $E_{inc}$ at frequency $\omega$. The response is investigated in the weak-excitation regime, and the atoms are modeled as point-like oscillators characterized by a polarizability $\alpha_j(\omega) = -(3\pi\varepsilon_0 \gamma_j c^3/\omega^3)(\omega - \widetilde{\Omega}_j)^{-1}$, $\varepsilon_0$ being the vacuum permittivity, $c$ the speed of light, $\gamma_j$ the natural linewidth of the atomic resonance and $\widetilde{\Omega}_j = Re(\widetilde{\Omega}_j) + i(n\gamma_j + \gamma_j^*)/2$ the complex transition frequency, with $\gamma_j^*$ a phenomenological non-radiative rate and $n$ the refractive index of the host material.[26] We adopt an $exp(i\omega t)$ notation. The oscillators are assumed to be linearly polarized, but the formalism can easily be developed for other polarizations. We denote by $p_j = p_j e_j$ the electric dipole moment of oscillator $j$, where $e_j$ ($|e_j| = 1$) defines the polarization orientation. The coupled-dipole system of equations is

$$p_j(r_j) = \alpha_j \{E_d(r_j) + \mu_0 \omega^2 \sum_{i \neq j} G_0(r_j, r_i; \omega) p_i + \mu_0 \omega^2 \sum_i \Delta G(r_j, r_i; \omega) p_i\} \cdot e_j, \tag{1}$$

where $E_d$ is the coherent driving field at frequency $\omega$, $G(r_j, r_i; \omega) = G_0(r_j, r_i; \omega) + \Delta G(r_j, r_i; \omega)$ is the dyadic Green's tensor written as a sum of the free-space Green's tensor

$$G_0(r_i, r_j; \omega) = \frac{\exp(-ikR_{ji})}{4\pi\epsilon_0 R_{ji}} \left[ \left(1 - \frac{ikR_{ji}+1}{k^2 R_{ji}^2}\right) I + \frac{3 + 3ikR_{ji} - k^2 R_{ji}^2}{k^2 R_{ji}^2} \frac{R_{ji} R_{ji}}{R_{ji}^2} \right], \tag{2}$$

and the regular scattering part $\Delta G(r_i, r_j; \omega)$, which accounts for light scattering by the resonator. In Eq. (2), $I$ is the 3×3 unit Dyad, $k = n\omega/c$, $R_{ji} = |R_{ji}| = |r_j - r_i|$, and $R_{ji} R_{ji}$ is the outer product of $R_{ji}$ with itself. The second term in Eq. (1) represents the direct dipole-dipole interaction between the oscillators. The first and third terms are more challenging to compute as they encompass the indirect interaction due to the scattering by the resonator. In our approach, we express the indirect interaction as a superposition of a few Quasi-Normal Modes (QNMs).[27] Close to a resonance, such expansions are able to predict the interaction with great efficiency[27-29] and a high accuracy, see Supplementary Material (SM) for details.

We first consider the driving field $E_d$. It is the sum of the background incident laser field $E_{inc}$ and the background field scattered by the resonator. Expanding the scattered contribution in the QNM basis, we have

$$E_d(r_j, \omega) = E_{inc}(r_j) + \sum_m \beta_m(\omega) \widetilde{E}_m(r_j), \tag{3}$$

where $\widetilde{E}_m(r_j)$ denotes the electric-field distribution of the $m^{th}$ QNM with a complex eigenfrequency $\widetilde{\omega}_m$ and $\beta_m(\omega) = \frac{\omega}{\widetilde{\omega}_m - \omega} \iiint \Delta\varepsilon(\omega, r) E_{inc}(\omega, r) \cdot \widetilde{E}_m(r) \, d^3r$ is the excitation coefficient,[30] expressed as an overlap integral between $E_{inc}$ and $\widetilde{E}_m$. In the previous expression, $\Delta\varepsilon(\omega, r)$ denotes the difference between the resonator and background permittivities. Similarly, we expand the last term in the brackets of Eq. (1), which represents the field scattered at $r_j$ by the resonator illuminated by the dipole $p_i$ located at $r_i$, as[27]

$$\mu_0 \omega^2 \Delta G(r_j, r_i; \omega) p_i = \sum_m \omega(\widetilde{\omega}_m - \omega)^{-1} (\widetilde{E}_m(r_i) \cdot p_i) \widetilde{E}_m(r_j). \tag{4}$$

Equation (1) can be recast in matrix form, which is convenient for numerical implementations. By denoting $\mathbf{P}$ the vector formed by the induced dipole modulus, $p_j$, $j = 1, \ldots N$, we obtain the master equation

$$\boldsymbol{\alpha} \boldsymbol{b} = (I - \boldsymbol{\alpha} A_{cav} - \boldsymbol{\alpha} B) \mathbf{P}. \tag{5}$$

In Eq. (5), $\boldsymbol{\alpha}$ is the diagonal $N \times N$ polarizability-matrix with $\alpha_{jj} = \alpha_j(\omega)$ and $\boldsymbol{b}$ is the $N$-vector formed from the driving laser field, $b_j = E_d(r_j) \cdot e_j$. The $N \times N$ dipole-dipole interaction matrix $B$, $B_{ji} = \mu_0 \omega^2 e_j \cdot G_0(r_j, r_i; \omega) e_i$ if $j \neq i$ and 0 otherwise, is crucial at small interatomic distances since it is responsible for a large collective Lamb shift[31] and subradiant emission.[32] The $N \times N$ resonator-interaction matrix, $\boldsymbol{A}_{cav} = \sum_m \omega(\widetilde{\omega}_m - \omega)^{-1} \boldsymbol{u}_m \boldsymbol{u}_m^T$, with $\boldsymbol{u}_m$ the column vector such that $u_{m,j} = \widetilde{\boldsymbol{E}}_m(\boldsymbol{r}_j) \cdot \boldsymbol{e}_j$, enables efficient long-range interactions via the resonance modes of the resonator, and is responsible for the formation of superradiant states,[9] even at low atomic densities.

In Eq. (5), every individual element of $\boldsymbol{b}, \boldsymbol{B}$ or $\boldsymbol{A}_{cav}$ is known analytically for all dipole locations and for any driving field distribution or frequency. Thus, once the few pertinent resonant QNMs are computed with full-wave calculations, one simply needs to inverse a $N \times N$ matrix to compute the induced dipoles.

**HYBRID EIGENSTATES**

The analyticity of the QNM expansion brings yet another considerable benefit: a direct computation of all sub/super-radiant states of the hybrid system. Without this expansion, purely numerical techniques, such as iterative pole-searching in the complex plane,[33] should be used. The eigenstates $\widetilde{\boldsymbol{P}}$ and the associated complex eigenfrequencies $\widetilde{\omega}$ would then be computed sequentially, one by one, and each pole computation would require a good initial guess value and a few full-Green tensor computations. The whole set of results reported hereafter would not have been obtained on a reasonable timescale. Conversely, with the present formalism, the $N$ eigenstates are easily and accurately computed all at once, simply as eigenvectors of a generalized eigenproblem. This greatly decreases the overall computation time, and makes exploring parameter space feasible.

The eigenstates $\widetilde{\boldsymbol{P}}$ are found by solving Eq. (5) in the absence of a driving field. In the SM, assuming that the oscillators are all identical ($\widetilde{\Omega}_j \equiv \widetilde{\Omega}$), we show that this equation can be naturally cast into a polynomial expansion in $\widetilde{\omega}$ by expanding the exponential term of $\boldsymbol{G}_0(\boldsymbol{r}_j, \boldsymbol{r}_i; \widetilde{\omega})$ in a Taylor series in $\Delta\widetilde{\omega} = \widetilde{\omega} - \widetilde{\Omega}$, $exp(in\widetilde{\omega}R_{ji}/c) = exp(in\widetilde{\Omega}R_{ji}/c)(1 + in\Delta\widetilde{\omega}R_{ji}/c + \cdots)$, which rapidly converges since $c/\widetilde{\Omega} \gg c/\Delta\widetilde{\omega}$ is broadly comparable to the resonator dimensions. Interestingly, if only the first two terms of the series are retained, the polynomial expansion becomes a generalized eigenvalue problem

$$\rho\left[\frac{\epsilon_0 c^3}{\widetilde{\Omega}^3}\boldsymbol{u}_m\boldsymbol{u}_m^T - \left(1 - \frac{\widetilde{\omega}_m}{\widetilde{\Omega}}\right)\boldsymbol{e}^T\boldsymbol{G}\boldsymbol{e}\right]\widetilde{\boldsymbol{P}} = \frac{\Delta\widetilde{\omega}}{\widetilde{\Omega}}\left[\left(1 - \frac{\widetilde{\omega}_m}{\widetilde{\Omega}}\right)\boldsymbol{I} + \rho\left(\frac{\epsilon_0 c^3}{\widetilde{\Omega}^3}\boldsymbol{u}_m\boldsymbol{u}_m^T + \boldsymbol{e}^T\boldsymbol{G}\boldsymbol{e} + \left(1 - \frac{\widetilde{\omega}_m}{\widetilde{\Omega}}\right)\boldsymbol{e}^T\boldsymbol{H}\boldsymbol{e}\right)\right]\widetilde{\boldsymbol{P}}. \tag{6}$$

The CPU time to solve Eq. (6) for $N = 100$ on a standard desktop computer is $\approx 1$ sec, which allows one to perform statistical analysis. In Eq. (6), we use the dimensionless coefficient $\rho = 3\pi\gamma_0/\widetilde{\Omega}$, and we use the fact that $\boldsymbol{e} \cdot \boldsymbol{G} \boldsymbol{e}$ and $\boldsymbol{e} \cdot \boldsymbol{H} \boldsymbol{e}$ are $N \times N$ matrices derived from the Taylor series defined by Eqs. (E5a) and (E5b) in the SM, where we additionally provide a simple program to solve Eq. (6). A single predominant QNM is assumed to derive Eq. (6), but an extension to multiple-resonances is straightforward, and importantly, does not change the nature of the numerical solution which remains a simple eigenvalue problem. Only when higher-order terms are retained in the Taylor series does Eq. (6) become a polynomial eigenvalue problem, but this can still be solved efficiently with modern eigensolver libraries. Tailoring Eq. (6) to deal with arbitrary polarizability tensors is also straightforward; it requires replacing the Green's-tensor projections on the polarization directions (symbolized by the $\boldsymbol{e}$ vectors in Eq. (6)) by more general projection matrices.

**RESULTS AND DISCUSSION**

In the following, we study hybrid systems composed of a large number of oscillators, and despite the complexity of the problem, we perform statistical computations and identify general collective properties that are robust to atomic disorder and should be generally observed in many experiments. As a resonator reference, we consider a gold nanorod (diameter $D = 30\ nm$, length $L = 100\ nm$) in silica ($n = 1.5$). The dipole-cavity interaction is dominantly mediated by the fundamental z-polarized electric-dipole resonance mode of the nanorod, see Figs. 1a or 1c. This mode with a complex frequency $2\pi c/\widetilde{\omega}_m = 920 + 47i\ nm$ is computed and appropriately normalized with an open-source code[30], which requires a few minutes per QNM on a standard desktop computer. We assume a Drude frequency-dependent permittivity of gold, $\varepsilon_{gold} = 1 - \omega_p^2/(\omega^2 - i\omega\Upsilon)$, with $\omega_p = 1.26\ 10^{16}\ s^{-1}$ and $\Upsilon = 1.41\ 10^{14}\ s^{-1}$. The molecules are considered as classical oscillators ($\widetilde{\Omega} = 2.05\ 10^{15} + i56.25\ 10^6\ Hz$) with a natural linewidth $\gamma_0 = 75$ MHz in vacuum and a resonant wavelength of 920 nm, matched with the plasmon dipolar resonance frequency. Non-radiative decays are not considered hereafter, $\gamma_j^* = 0$.

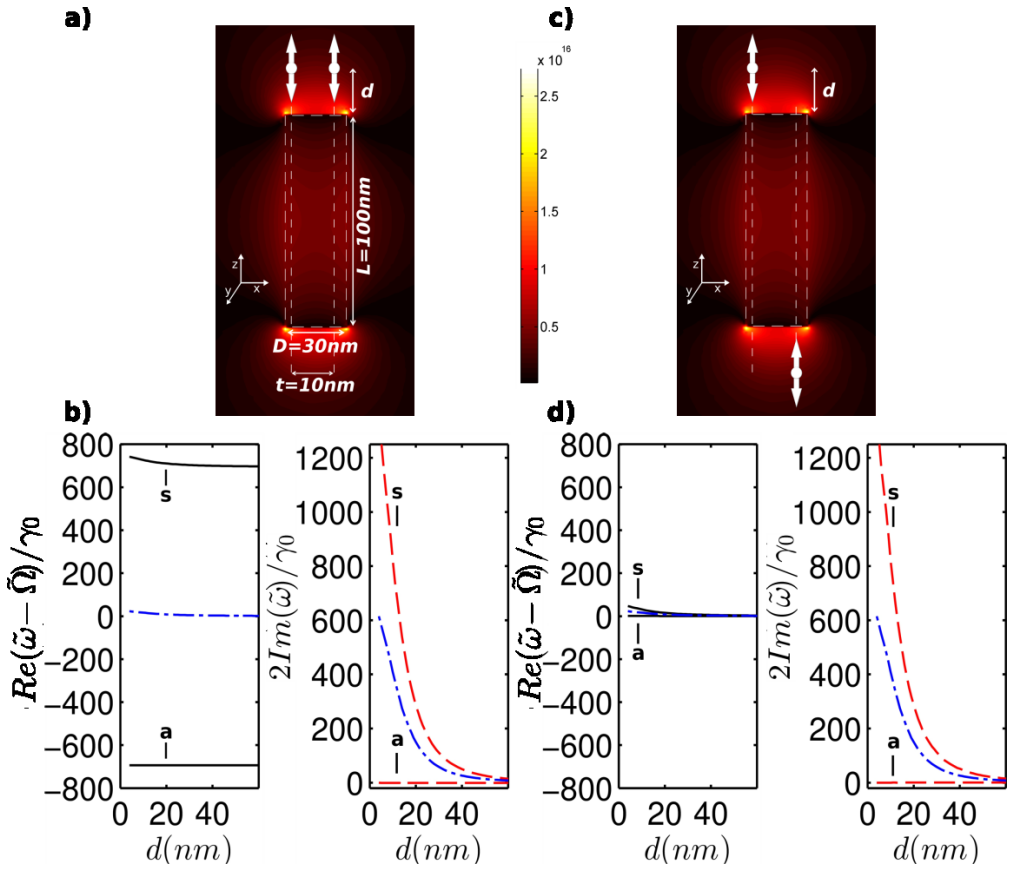

**Figure 1. Subradiant ("$a$") and superradiant ("$s$") states of molecular doublets.** The two dipoles, shown with white arrows in **a**) and **c**), are oriented parallel to the gold nanorod (diameter $D$ = 30 nm, length $L$ = 100 nm) and located at a separation distance $d$ from the nanorod surface. In **a**) and **b**) the dipoles are on the same pole of the rod, while in **c**) and **d**) they are on opposite poles. The separation distance between the dipoles is denoted by $t$. **b**) and **d**) $Re(\widetilde{\omega} - \widetilde{\Omega})$ and $2Im(\widetilde{\omega})$ as a function of $d$. The blue dash-dotted curves hold for a single oscillator placed on the rod axis and oriented parallel to the rod. The normalized electric-field distribution $|\widetilde{E}_z|$ in $Vm^{-1}$ of the electric-dipole resonance mode ($2\pi c/\widetilde{\omega}_m = 920 + 47i\ nm$) is shown in **a**) or **c**).

**Hybrids with two molecules.** Before analyzing large ensembles, it is instructive to consider parallel doublets placed in the bright near-field spot at a distance $d$ from the rod. To distinguish the impact of "Lamb-shift" and Purcell effects on the doublet properties, we consider two distinct cases (Figs. 1a and 1c). The oscillators are either placed on the same side, or on opposite sides of the rod, and at the same distance from the rod. The interaction of the oscillators with the localized plasmon is the same in both cases, but the direct dipole-dipole interaction considerably changes. For each case, the doublet eigenstates are formed by asymmetric (dark/subradiant) and symmetric (bright/superradiant) states, respectively denoted by "$a$" and "$s$".

Figures 1b and 1d show the eigenfrequencies $\widetilde{\omega}$. Here, $\Delta = Re(\widetilde{\omega} - \widetilde{\Omega})$ and $2Im(\widetilde{\omega})$ represent the frequency shift and linewidth of the doublet. These figures convey some important information that will help analyze hybrids with many oscillators. First, the shift is nearly independent of $d$; rather, it depends on the separation distance $t$ between the oscillators, increasing as $(kt)^{-3}$ [34] for close oscillators. Second, the linewidths of the "$a$" states remain very small for all distances $d$, consistent with the fact that the dark states do not couple to the localized electric-dipole mode for symmetry reasons. Note that, in reality, quenching should become dominant at small distances, but this cannot be modelled with our single-QNM-expansion approximation. Third, the linewidths of the "$s$" states are strongly enhanced for small values of $d$, becoming as large as $1000\gamma_0$ for $d < 10\ nm$. The decay-rate acceleration that preferentially impacts "$s$" modes is the analogue for doublets of the traditional nanoantenna effect.[35] The latter is obtained when the light emission properties of single quantum emitters are profoundly altered by antennas, increasing optical excitation rates, modifying radiative and nonradiative decay rates or emission directionality. The acceleration is about twice that of the same single dipole placed and oriented along the rod axis (blue dash-dotted curves in Figs. 1b and 1d). Similar accelerations will be also observed for large ensembles.

**Hybrids with many molecules.** Figure 2a shows the scattering ($\sigma_{sca}$) and absorption ($\sigma_{abs}$) cross-section spectra of a hybrid system with $N = 100$ oscillators under illumination by a plane-wave polarized along the z-axis of the rod. The oscillators are randomly polarized and positioned within a 15-nm-thick shell located at a 15-nm minimum distance from the rod. Quenching is thus negligible. The minimum distance $t_{min}$ between neighboring dipoles is 10 nm. Both spectra exhibit a series of sharp peaks superimposed on a spectrally-broad ($\sim 10^3 \gamma_0$) background dip. Note that the absorption and scattering cross-sections of the bare nanorod (without oscillators) are nearly constant ($\approx 0.04\ \mu m^{-2}$) in the spectral range of interest.

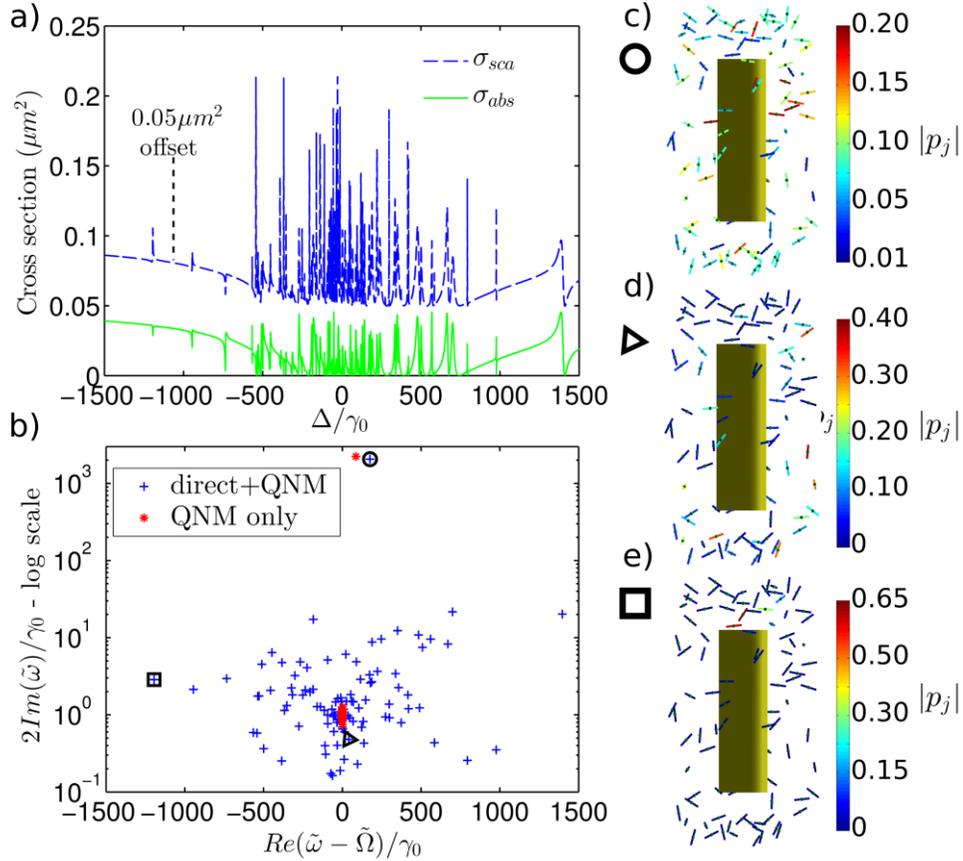

**Figure 2. Hybrids with many oscillators ($N$ = 100). a)** Scattering and absorption cross sections of the hybrid under illumination by a plane wave polarized along the rod axis. **b)** Eigenstate energies and decay rates (blue pluses) obtained by computing the eigenvalues of Eq. (6). The red circles are obtained by neglecting the direct dipole-dipole interaction. **c)-e)** Schematic representation of the dipole-moment distributions of 3 normalized states indicated by the marks o (superradiant), ▷, □ in **b**). Every dipole $j$ is colored according to its normalized dipole moment value $|p_j|$.

To improve our understanding, we compute the hybrid eigenfrequencies. The latter are displayed in Fig. 2b with blue crosses. We find a single superradiant state and a myriad of subradiant states. Careful inspection shows a one-to-one correspondence between the sharp peaks in Fig. 2a and the eigenfrequencies. Hereafter, we will conclusively show that the broad background dips in Fig. 2a are due to the superradiant state. Figures 2c-2e represent the dipole-moment distributions of 3 states, as indicated by the marks o, ▷, □ in Fig. 2a, where every dipole is colored according to its normalized dipole moment value $|p_j|$, $\sum_j |p_j|^2 = 1$. Almost half of the dipoles of the superradiant state have a significant moment, revealing a strong cooperativity over a large spatial extent. We estimate that 70% (see SM) of the subradiant states are also extended, involving up to 20% of the dipoles (Fig. 2d), while the other remaining 30% are localized states formed by the coupling of 2-5 dipoles (Fig. 2e).

We have repeated the eigenstate computations, neglecting the direct interaction due to the free-space Green's function. The new eigenfrequencies are shown with red dots in Fig. 2b. The superradiant state is weakly impacted. The significant result is the marked deviation for the cloud of subradiant states that becomes degenerate; we have verified that their cooperativity becomes large, with 15% to 35% of active dipoles.

**Ensemble-averaged properties.** In general, even when considering single quantum hybrids, the exact positions and orientation of the atoms or molecules are not precisely known. Also one may be interested in controlling the optical macroscopic properties of hybrid ensembles, rather than individual ones.[36-37] Hence, the effective properties that are robust to ensemble averaging are essential for further investigations.

We have performed extensive computations, randomly varying the oscillator polarizations and positions. Figure 3 summarizes the main relevant results obtained by averaging over $10^4$ random configurations. The averaged extinction spectrum (Fig. 3a) exhibits a smooth shape. We ascertain that the prevalent feature—the presence of an extinction dip that broadens as $N$ gradually increases—is a direct consequence of superradiant states. These states withstand ensemble averaging and essentially behave as super oscillators[4] with very large polarizabilities that efficiently dull the field at the resonator. Thus, the scattering and absorption cross-sections both diminish on resonance, yielding a minimum extinction as small as 0.02 µm$^{-2}$—a value 4-times smaller than the extinction of the bare nanorod. Strikingly, the oscillator shell acts as a type of "invisibility cloak" with a bandwidth that increases with $N$. In contrast, the sharp peaks due to the subradiant states are largely smoothed out by averaging. As inferred from other computations, we believe that the subradiant states mostly contribute to increasing scattering and absorption for frequencies just around the oscillator frequency and are responsible for the "camel hump" shape of the broad dip.

To unambiguously demonstrate that the broad dip is a remnant of a coherent collective effect, we compute the average decay rate $\langle \Gamma_{sup} \rangle$ of the superradiant states as a function of $N$. The results are shown in Fig. 3b for two distinct values of $t_{min}$. As before, we repeat the same computations by neglecting the dipole-dipole interaction; the new data (red triangles) now scale perfectly with $N$, and are well fitted with an intrinsic characteristic of the hybrid:

$$\langle \Gamma_{\text{sup}} \rangle = N\gamma_0 \langle F_p \rangle, \tag{7}$$

which is the product of $N\gamma_0$ [4] with the mean Purcell Factor $\langle F_p \rangle$ experienced on average by the randomly-positioned oscillators. It then becomes clear that the direct dipole-dipole interaction alters the cooperativity[38] at large densities, preventing the realization of ultra-bright hybrid states by combining the Dicke and Purcell effects in a perfect way. This result, which is somewhat in contradiction with Ref. [39], would be interesting to investigate with geometries exhibiting a larger number of oscillators in remote positions, for instance microcavities with larger mode volumes and larger Q's.

The resonator is not just enhancing the superradiant decay rate by a factor $\langle F_p \rangle$, as a naïve reading of Eq. (7) might indicate. In fact, the primarily impact of the resonator is to reinforce long-range interaction, so that $\Gamma_{\text{sup}}$ actually scales with $N$, even over large volumes. For instance, the superradiant decay rate in Fig. 2b is $2060\gamma_0 \approx 100\gamma_0 \langle F_p \rangle$ with $\langle F_p \rangle = 23.3$, whereas it is only $3.4\gamma_0 \ll 100\gamma_0$ for the same oscillator distribution in a bulk medium without a nanorod. In contrast to subradiant states, superradiant states are artificially reinforced by the resonator, irrespective of the particular oscillator position and polarization orientation. Cooperativity is resistant to spatial disorder and the averaged responses, such as the extinction dip in Fig. 3, are actually general features that can be expected for any hybrids. This essential property was not discussed in previous studies restricted to the analysis of ensembles with specific symmetries and polarization orientations.[9,15]

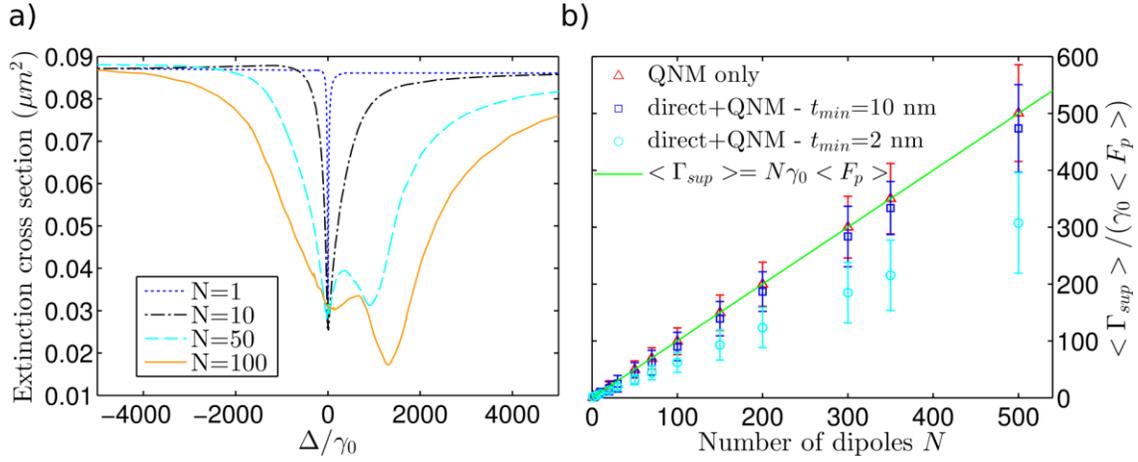

**Figure 3. Position/orientation averaged properties of hybrids for large ensembles. a)** Averaged extinction spectra of nanohybrids composed of *N* = 1, 10, 50 and 100 oscillators randomly distributed and oriented around the nanorod. **b)** Evolution of the average decay rate $\langle \Gamma_{sup} \rangle$ of the superradiant state. Equation (7) (green line) is perfectly verified if one assumes that the electromagnetic interaction is only mediated by the plasmons, neglecting the direct dipole-dipole interaction. The error bars indicate statistical errors.

## CONCLUSION

In conclusion, we introduced a modal formalism to study the response of large ensembles of atoms or molecules coupled via electromagnetic resonances. The use of a modal-expansion brings analyticity in the treatment and for the first time allows one to model complex hybrid systems without any restriction on the resonator shapes, material properties, nor polarization orientations.

Some essential properties of hybrids, such as the extinction dip in Fig. 3, which are robust to spatial and polarization disorders, are likely to be observed in many experiments on single hybrids or disordered collections of hybrid metamaterials.[36-37] We have shown that they are dominantly driven by superradiant states, whose decay rates scale with the number of dipoles[4] and an intrinsic characteristic of the resonator, the spatially-averaged Purcell Factor.

The modal formalism was derived for linearly-polarized oscillators, such as molecules, but it is straightforward to extend this approach to encompass other polarizations, as required in typical experiments with quantum dots,[25] carbon nanotubes,[37] or cold atoms.[8] Finally, let us mention that many quantum formalisms rely on a treatment of the emission field with a classical dyadic Green's-tensor formalism.[15] Thus the present 'toolkit' represents an important step toward advanced formalisms to model the evolution dynamics of hybrids beyond the perturbation regime.

**Acknowledgements.** Spyridon G. Kosionis and Pierre Fauché acknowledge the LabEx LAPHIA for post-doctoral and Ph. D fellowship as well as financial support. The authors thank K. Vynck, R. Vallée and B. Lounis for fruitful discussions, and B. Barrett for careful proofreading.

# Supplementary Material

# Collective scattering in hybrid nanostructures with many atomic oscillators coupled to an electromagnetic resonance

Pierre Fauché,[1,2] Spyridon G. Kosionis[1] and Philippe Lalanne[1,*]

[1] Laboratoire Photonique, Numérique et Nanosciences (LP2N), UMR 5298, CNRS-IOGS-Univ Bordeaux, Institut d'Optique d'Aquitaine, 33400 Talence, France.
[2] CNRS, Univ Bordeaux, CRPP, UPR 8641, 115 Avenue Schweitzer, 33600 Pessac, France.

* Corresponding author: philippe.lalanne@institutoptique.fr

## Content

## 1. Test of the accuracy of the QNM expansion formalism

In Fig. 2 of main text, we analyze the property of the dark and bright states of a molecule doublet as a function of the separation distance $d$ between the molecules and the nanorod.

In Fig. S1, we show the scattering and absorption cross sections of the hybrid under illumination by a plane-wave incident normally to the cross-section plane shown in Figs. 2a and 2c, and polarized along the rod z-axis. The results are obtained for molecules at $d = 15$ nm from the rod surface. Note that the molecule dimers have been transversely shifted by 1 nm towards the positive x-direction so to break symmetries that would prevent exciting the dark subradiant state.

The broad dips corresponds to the excitation of the superradiant state "s", while the narrow resonance corresponds to the subradiant state "a".

Strikingly, both the scattering and absorption cross-sections vanish on resonance, yielding an extinction dip for the superradiant state as small as 0.001 $\mu m^{-2}$, a value 80-times smaller than the absorption and scattering cross-sections of the isolated nanorod (black lines in Fig. S1). This phenomena has been previously predicted for single quantum emitters placed in the mouth of two metallic nanocones [Che13] and has been shown to be a general property of hybrids formed by a single molecule coupled to metallic nanoantennas, provided that the molecule is subject to a large Purcell effect [Yan15]. Since the superradiant doublet essentially behaves as a super molecule with a polarizability twice larger, unsurprisingly we find that the doublet fully cloaks the nanorod on resonance, leading to full transparency for the incoming light beam.

Two series of results are shown in Fig. S1:
- The circles are obtained by solving the master Eq. (5) in the main text using COMSOL Multiphysics®, implying that for every frequency, the driving field on each molecule is computed exactly by solving the scattering of a plane plane by the nanorod and that the full-wave Green tensor of every molecule is also solved numerically using COMSOL.

- The solid and dashed curves are obtained by using the QNM expansion, retaining a single QNM in the expansion, the fundamental electric dipole mode. The QNM computation and normalization are performed using the method presented in [Bai13]. The computation is very fast and lasts less than two minutes without using any symmetry on a standard computer.

We notice that the general asymmetric response, including the deep minima and the narrow resonance, are well predicted with the QNM-expansion approach for both the scattering and absorption cross sections. The only apparent discrepancy occurs for the absorption cross-section on the red side of the broad resonance, but this slight deviation has already been observed in earlier works on QNM expansion [Bai13,Sau13] and it is thus not a direct consequence of the present multi-oscillator formalism.

In our opinion, the overall excellent agreement between the two series of data in Fig. S1 provides a strong support for the accuracy of the present QNM formalism.

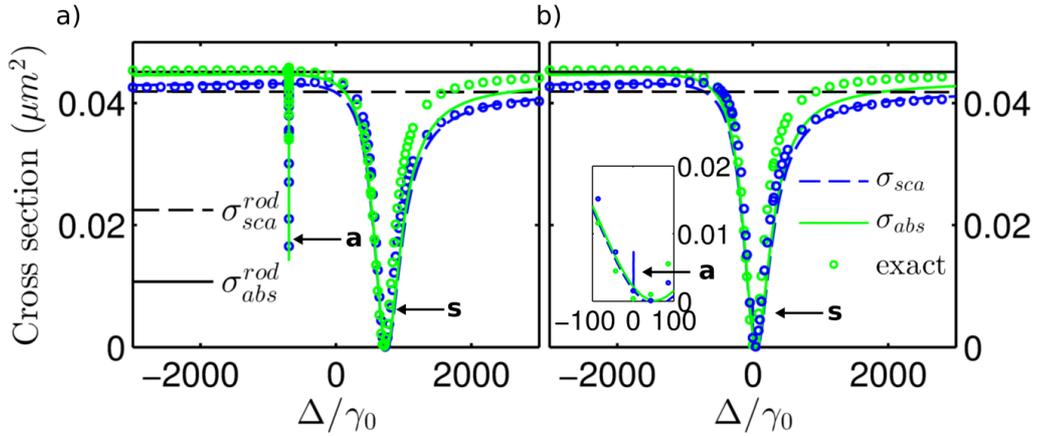

**Figure S1** Scattering and absorption cross sections of the molecular doublets shown in Fig. 2 in the main text as a function of the normalized frequency detuning $\Delta/\gamma_0$ around 920 nm. The molecules are located either on the same pole of the nanorod (a) or on the opposite poles (b). The scattering and absorption cross-sections are shown in blue and green, respectively. Data obtained with the QNM expansion are shown with solid-green and blue-dashed curves. Circles are fully-vectorial computational results obtained with COMSOL. The scattering and absorption cross sections of the bare nanorod, without molecule, are shown for comparison with the black lines. The resonance due to the symmetric (asymmetric) state is designated with "s" ("a"). The inset in b) shows a zoom around $\Delta = 0$ and reveals the subradiant resonance.

## 2. Computation of the scattering and absorption cross-sections

From the knowledge of the molecule dipole moments $p_j$, one may calculate the extinction cross-section $\sigma_{ext}$ with the present formalism by using the following formula

$$\sigma_{ext} = -\frac{\omega}{2S_0}\sum_{j=1\rightarrow N} Im[\boldsymbol{p}_j \cdot \boldsymbol{E}_{inc}(\boldsymbol{r}_j,\omega)] - \frac{\omega}{2S_0}\iiint Im[\Delta\epsilon(\boldsymbol{r},\omega)(|\boldsymbol{E}_{inc}(\boldsymbol{r},\omega)|^2 + \boldsymbol{E}_{sca}^{tot}(\boldsymbol{r},\omega) \cdot \boldsymbol{E}_{inc}^*(\boldsymbol{r},\omega))]d^3\boldsymbol{r}, \quad (S1)$$

where the first term describes the extinction cross section due to the molecules and the second term accounts for the nanorod. In Eq. S1, $S_0$ is the time-averaged Poynting vector of the incident field $\boldsymbol{E}_{inc}$ and $\boldsymbol{E}_{sca}^{tot}$ denotes the total scattered field

$$\boldsymbol{E}_{sca}^{tot}(\boldsymbol{r},\omega) = \left(\beta_m(\omega) + \sum_{j=1\rightarrow N} \omega(\widetilde{\omega}_m - \omega)^{-1}\widetilde{\boldsymbol{E}}_m(\boldsymbol{r}_j) \cdot \boldsymbol{p}_j\right)\widetilde{\boldsymbol{E}}_m(\boldsymbol{r}). \quad (S2)$$

Then, assuming absorptionless molecules, the scattering cross section $\sigma_{abs}$ is given by

$$\sigma_{abs} = -\frac{\omega}{2S_0} \iiint Im[\epsilon(\boldsymbol{r},\omega)] |\boldsymbol{E}_{inc}(\boldsymbol{r},\omega) + \mu_0 \omega^2 \sum_{j=1\to N} \boldsymbol{G}_0(\boldsymbol{r},\boldsymbol{r}_j,\omega)\boldsymbol{p}_j + \boldsymbol{E}_{sca}^{tot}(\boldsymbol{r},\omega)|^2 d^3\boldsymbol{r}, \quad \text{(S3)}$$

where $\epsilon(\boldsymbol{r},\omega)$ is the resonator permittivity.

## 3. Derivation of the generalized eigenvalue problem

We denote by $\widetilde{\omega} = \widetilde{\Omega} + \Delta\widetilde{\omega}$ the eigenfrequency of the eigenstate $\widetilde{\boldsymbol{P}}$, and introduce the following notations for the sake of simplicity

$$\Delta\widetilde{\omega}/c = \Delta k, \widetilde{\Omega}/c = K \text{ and } \widetilde{\omega}_m/c = K_m, \quad \text{(E1)}$$

with $\Delta\widetilde{\omega}$ the complex frequency shift, $\widetilde{\Omega}$ the complex transition frequency of the molecules and $\widetilde{\omega}_m$ the resonance frequency of the resonator. Thus the molecule polarizability simply reads as $\alpha(\omega) = \frac{-3\pi\varepsilon_0\gamma_0/c}{\Delta k(\Delta k+K)^3}$.

The eigenstates are solutions of the master equation, see Eq. (5) in the main text, without external driving field ($\boldsymbol{b} = 0$)

$$(\boldsymbol{I} - \boldsymbol{\alpha}\,\boldsymbol{A}_{cav} - \boldsymbol{\alpha}\,\boldsymbol{B})\,\widetilde{\boldsymbol{P}} = 0, \quad \text{(E2)}$$

with, according to the main text,

$$\boldsymbol{\alpha}\boldsymbol{A}_{cav} = \frac{3\pi\varepsilon_0\gamma_0/c}{\Delta k(\Delta k+K-K_m)(\Delta k+K)^2}\boldsymbol{u}_m \boldsymbol{u}_m^T, \quad \text{(E3a)}$$

$$(\boldsymbol{\alpha}\,\boldsymbol{B})_{ji} = \frac{-3\pi\gamma_0/c}{\Delta k(\Delta k+K)}\boldsymbol{e}_j \cdot \boldsymbol{G}_0(\boldsymbol{r}_j,\boldsymbol{r}_i;\widetilde{\omega})\boldsymbol{e}_i, \quad \text{(E3b)}$$

if $j \neq i$ and 0 otherwise. The analytic expression of the free-space Green tensor $\boldsymbol{G}_0(\boldsymbol{r}_j,\boldsymbol{r}_i;\widetilde{\omega})$ at complex frequency $\widetilde{\omega}$ is given by Eq. (2) in the main text. Since the subradiant and superradiant states all occur in a narrow energy interval centered about the complex transition frequency $\widetilde{\Omega}$, $\boldsymbol{G}_0(\boldsymbol{r}_j,\boldsymbol{r}_i;\widetilde{\omega})$ can be expanded as a Taylor series in $\Delta k$ at $\widetilde{\Omega}$

$$\boldsymbol{G}_0(\boldsymbol{r}_j,\boldsymbol{r}_i;\widetilde{\omega}) = \boldsymbol{G}(\boldsymbol{r}_j,\boldsymbol{r}_i) + \frac{\Delta k}{K}\boldsymbol{H}(\boldsymbol{r}_j,\boldsymbol{r}_i) + O\left(\frac{\Delta k^2}{K^2}\right). \quad \text{(E4)}$$

With the notation $v_{ji} = nKR_{ji}$ ($R_{ji}$ is the distance between molecules $i$ and $j$), elementary algebra lead to

$$G_{ji} = \boldsymbol{e}_j^T \boldsymbol{G}(\boldsymbol{r}_j,\boldsymbol{r}_i)\boldsymbol{e}_i = \frac{exp(-iv_{ji})}{4\pi v_{ji}/n}\boldsymbol{e}_j^T \cdot \left[\widehat{\boldsymbol{I}} - \frac{\boldsymbol{R}_{ji}\boldsymbol{R}_{ji}}{R_{ji}^2} - v_{ji}^{-2}(1+iv_{ji})\left(\widehat{\boldsymbol{I}} - \frac{3\boldsymbol{R}_{ji}\boldsymbol{R}_{ji}}{R_{ji}^2}\right)\right]\boldsymbol{e}_i, \quad \text{(E5a)}$$

$$H_{ji} = \boldsymbol{e}_j^T \boldsymbol{H}(\boldsymbol{r}_j,\boldsymbol{r}_i)\boldsymbol{e}_i = \frac{exp(-iv_{ji})}{4\pi iv_{ji}^2/n}\boldsymbol{e}_j^T \cdot \left(\frac{3\boldsymbol{R}_{ji}\boldsymbol{R}_{ji}}{R_{ji}^2} - \widehat{\boldsymbol{I}}\right)\boldsymbol{e}_i - iv_{ji}G_{ji}, \quad \text{(E5b)}$$

which can be found in the main text.

We now again consider Eq. (E2). Multiplying all terms by $\Delta k(\Delta k + K - K_m)(\Delta k + K)^2$, we obtain

$$\Delta k(\Delta k + K - K_m)(\Delta k + K)\,\widetilde{\boldsymbol{P}} +$$

$$\frac{3\pi\gamma_0}{c}\left(\frac{-\varepsilon_0}{\Delta k+K}\boldsymbol{u}_m\boldsymbol{u}_m^T + K(\Delta k + K - K_m)\left(\boldsymbol{e}^T\boldsymbol{G}\boldsymbol{e} + \frac{\Delta k}{K}\boldsymbol{e}^T\boldsymbol{H}\boldsymbol{e} + O\left(\frac{\Delta k^2}{K^2}\right)\right)\right)\widetilde{\boldsymbol{P}} = 0. \quad \text{(E6)}$$

This equation shows that, if only the first two terms are retained in the Taylor series of $\boldsymbol{G}_0(\boldsymbol{r}_j,\boldsymbol{r}_i;\widetilde{\omega})$, only terms in $O(k/K)$ have to be considered in all terms of Eq. (E6). Thus Eq. (E6) becomes

$$\frac{\Delta k}{K}\boldsymbol{B}\widetilde{\boldsymbol{P}} = \boldsymbol{A}\widetilde{\boldsymbol{P}}. \quad \text{(E7)}$$

with

$$\boldsymbol{A} = \rho\left[\frac{\varepsilon_0}{K^3}\boldsymbol{u}_m\boldsymbol{u}_m^T - \left(1 - \frac{K_m}{K}\right)\boldsymbol{e}\cdot\boldsymbol{G}\boldsymbol{e}\right], \quad \text{(E8)}$$

$$B = \left(1 - \frac{K_m}{K}\right)I + \rho\left(\frac{\varepsilon_0}{K^3}u_m u_m^T + e^T G e + \left(1 - \frac{K_m}{K}\right)e^T H e\right), \tag{E9}$$

where the new dimensionless coefficient is $\rho$ defined by $\rho = 3\pi\gamma_0/\widetilde{\Omega}$.

Equation (E7) corresponds to a generalized eigenproblem. If higher-order terms have to be considered in the Taylor series of $G_0(r_j, r_i; \omega)$, Eq. (E7) is transformed into a polynomial eigenproblem that can still be solved very efficiently with modern eigensolver libraries.

The numerical implementation of the set of equations (E7-E9) is straightforward, as shown in the next Section.

## 4. Program to solve the eigenproblem

The following function, written with notations directly related to the equations of the previous Section, can be run in a Matlab environment. We just need to specify the number of molecules *N*, their positions **X**, **Y**, **Z**, their orientations and the QNM electric field at all molecule positions.

The International System of Units is used, but most variable are dimensionless.

```
function [complex_frequency, eigenstate]=find_pole(N,X,Y,Z,dipole_orientation, ...
ld_A,gamma_0,omega_pole,n_ext,Ex_mode_AtMolecule,Ey_mode_AtMolecule,Ez_mode_AtMolecule)
% 2016
% On a standard desktop computer, CPU time is about 1 sec for N=100 emitters
% Convention exp(iwt), SI units

%% Inputs :
% N: real number : number of molecules
% X : N × 1 vector : X-coordinate of the molecules
% Y : N × 1 vector : Y-coordinate of the molecules
% Z : N × 1 vector : Z-coordinate of the molecules
% D2: N × 3 matrix : dipole orientation of molecules (ex,ey,ez) denoted by e in Eqs. (E8-E9)
% ld_A : real number : molecule resonance wavelength, in meter
% gamma_0 : real number : molecule natural linewidth in a vacuum
% omega_pole : complex number : pole of the QNM
% n_ext : real positive number : refractive index of background medium
% Ex_mode_AtMolecule :  N × 1 vector : X-field at molecule locations
% Ey_mode_AtMolecule :  N × 1 vector : Y-field at molecule locations
% Ez_mode_AtMolecule :  N × 1 vector : Z-field at molecule locations

%% Outputs :
% complex_frequency: N × 1 vector : complex frequencies of all eigenstates
% eigenstate : N × N matrix : all eigenstates

%% Constants and initialization
c=299792458; % light speed
eps0=1/(4*pi*9e9); % vacuum permittivity
omega_molecule=2*pi*c/ld_A + 1i*gamma_0/2; % complex resonance frequency of the molecule (Ω̃)

%% computation of matrix A and B with a first order expansion of exp(iknR)
K=(omega_molecule)/c; % wavevector of the molecule in m-1
Km=omega_pole/c; % wavevector of the resonance mode in m-1

% dipole-QNM interaction
a=zeros(N_molecule,1);
for plo=1:N_molecule
   a(plo)=D2(plo,1:3)*[Ex_mode_AtMolecule(plo);Ey_mode_AtMolecule(plo);Ez_mode_AtMolecule(plo)];
end
AA=a*a.'*eps0/K^3;

% dipole-dipole interaction
G=zeros(N_molecule);H=G;
for plo1=1:N_molecule
   for plo2=1:N_molecule
      if plo1~=plo2
```

```
      R=[X(plo1)-X(plo2),Y(plo1)-Y(plo2),Z(plo1)-Z(plo2)]
      Xx=R.'*R/norm(R)^2;
      v=n_ext*K*norm(R); % v is defined just after Eq. (E4)
      R1=D2(plo2,1:3)*(eye(3)-Xx+(eye(3)-3*Xx)*(-i/v-1/v^2))*D2(plo1,1:3).';
      R2=-i/v*D2(plo2,1:3)*(3*Xx-eye(3))*D2(plo1,1:3).'-i*v*R1;
      G(plo1,plo2)=R1*exp(-i*v)/(4*pi*v/n_ext); % Eq. (E5a)
      H(plo1,plo2)=R2*exp(-i*v)/(4*pi*v/n_ext); % Eq. (E5b)
    end
  end
end

% Eigenstates computation with the generalized eigenproblem
rho=3*pi*(gamma_0/n_ext)/c/K; % dimensionless coefficient defined after Eq. (9)
B=(1-Km/K)*eye(N_molecule)+rho*(AA+G+H*(1-Km/K)); % see Eq. (E9)
A=rho*(AA-G*(1-Km/K)); % see Eq. (E8)
[lambda, eigenstate]=eig(A,B); % solve Av = lambda*Bv  the eigenvalue is Δk/K, see Eq. (E7)
complex_frequency=lambda*K*c+omega_molecule; % complex frequency of all eigenstates
```

## 5. Eigenstates of large ensembles: participation number

Not much has been said so far on the number of molecules involved into the formation of the subradiant and superradiant states. To investigate cooperativity effects in the formation the eigenstates in random hybrids, we consider the same gold nanorod (diameter $D$ = 30 nm, length $L$ = 100 nm) and dress it with 100 dipoles, randomly distributed and oriented. We compute all the eigenstates with the program of Section 3 and normalize each eigenstate such that

$$\sum_{j=1}^{N}|\tilde{P}_j|^2 = 1. \tag{E10}$$

To quantify the cooperativity, we use the participation number [Weg80]

$$p = \left(\sum_{j=1}^{N}|\tilde{P}_j|^4\right)^{-1}. \tag{E11}$$

This number, $1 \leq p \leq N$ not only estimates the collectivity of the eigenstate, i.e. the average number of active dipoles in the eigenstate, but also its spatial extension [Kra93]. Figure S2 shows the 100 eigenvalues in the complex frequency plane. Every eigenvalue is colored to represent the participation number according to a scale shown with a vertical bar on the right side of the figure.

We observe that 31% of the subradiant modes corresponds to localized modes, involving only two to five neighbour dipoles that are interacting directly or via the resonator. The results also clearly evidence that the subradiant states with large positive or negative frequency shifts (appearing in dark-blue) are localized modes with small $p$'s, confirming that direct dipole-dipole interactions are frequently involved in localized eigenstates. However we note that some localized subradiant states with small frequency shifts also exit; for specific relative positions and orientations of the molecules involved in the state, the direct-direct dipole interaction is weak.

The remaining 69% of the subradiant modes are extended collective modes, involving up to 20% of the dipoles, with small frequency shifts. Still, none of the subradiant is as collective/extended as the superradiant state. The largest participation number is reached by the superradiant mode where almost half of the molecule ($p = 47$) are involved.

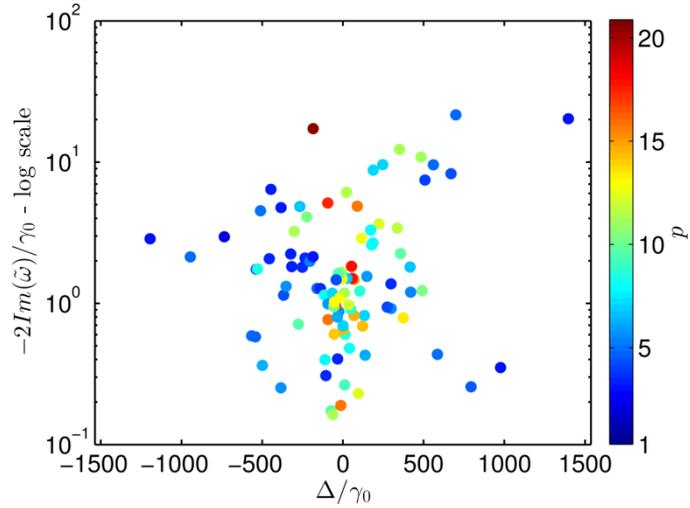

**Figure S2.** Cooperativity of the subradiant states (the superradiant state with $p = 47$ is not shown) of a complex hybrid formed with 100 molecules randomly positioned around a gold nanorod (diameter $D$ = 30 nm, length $L$ = 100 nm) embedded in a host medium of refractive index 1.5.